# Nanoscale Electric Field Sensing Using Levitated Nano-resonator with Net Charge

Shaocong Zhu[1], Zhenhai Fu[1,*], Xiaowen Gao[1,*], Cuihong Li[1], Zhiming Chen[1], Yingying Wang[1], Xingfan Chen[1], and Huizhu Hu[1,2,*]

[1]*Quantum Sensing Center, Zhejiang Lab, Hangzhou 310000, China*
[2]*State Key Laboratory of Modern Optical Instruments, College of Optical Science and Engineering, Zhejiang University, Hangzhou 310027, China*
*\*fuzhenhai@zju.edu.cn, \*gaoxw@zhejianglab.com, \*huhuizhu2000@zju.edu.cn*

**Abstract:** Nanomechanical resonator based on levitated particle exhibits unique advantages in the development of ultrasensitive electric field detector. We demonstrate a three-dimensional, high-sensitivity electric field measurement technology using the optically levitated nanoparticle with a known net charge. By changing the relative position between nanoparticle and parallel electrodes, the three-dimensional electric field distribution is scanned. The measured noise equivalent electric intensity with charge amount of 100 reaches the order of $1\mu V/cm/Hz^{1/2}$ at $1.4\times10^{-7}$mbar. Linearity analysis near resonance frequency shows a measured linear range over 91dB limited only by the maximum output voltage of the driving equipment. This work may provide avenue for developing a high-sensitive electric field sensor based on optically levitated nano-resonator.

1. Introduction

The ability to characterize static and time-dependent electric fields *in situ* with high sensitivity and high spatial resolution has profound applications for both fundamental science and technology. Precision sensing of electric fields and forces that couple to charge is the most direct way to search for deviations from Coulomb's law, which may be motivated by the presence of new forces under which dark matter could be charged [1,2]. Recent theoretical models pointed out that such new forces can weakly mix with electromagnetism, resulting in new Coulomb-like interactions [3]. Traditional electric field sensors to date mainly include the dipole antenna-coupled electronics [4], the electro-optic crystals [5,6,7], the resonant MEMS structures [8,9,10]. In addition, the recently emerging Rydberg atom-based sensors have demonstrated the capabilities of electric field distribution measurement with submillimeter spatial resolution [11,12] and the highest sensitivity up to order of $1\mu V/m/Hz^{1/2}$ ever reported from radio frequency to microwave electric fields [13,14,15].

The levitated nanomechanical resonator exhibits unique advantages in the development of precise force [16,17,18] and acceleration sensors [19,20] at the micro- and nanoscale, attributed to its high sensitivity and potential integration [21]. The nanomechanical system optically levitates the charged dielectric nanoparticle in high vacuum, thus making it a harmonic oscillator sensitive to the surrounding electric field. In case of a weak electric field, the harmonically driven response in power spectral density of oscillator's displacement is directly proportional to the electric intensity at its location and the amount of net charge it carries. Therefore, on the premise of knowing the amount of net charge, ultra-high force detection sensitivity means ultra-high electric field detection sensitivity.

In the present study, we extend previous works on highly sensitive force detection using optically levitated nano-resonator [22] to a novel, three-dimensional, high-sensitivity electric field measurement technology. Using the parallel plate electrodes as source of the electric field with known frequency, motion signal of the nanoparticle in the three orthogonal direction are used to measure the electric field vectors of corresponding axis. By changing the relative positions of the nanoparticle and the electrodes, electric field of the electrodes is scanned point-by-point, and the three-dimensional electric field mapping ability of the scheme is demonstrated.

By applying parametric feedback at 1.4×10$^{-7}$mbar, the force and electric intensity detection sensitivity equivalent from the measured displacement spectral density reach the order of 10$^{-20}$N/Hz$^{1/2}$ and 1μV/cm/Hz$^{1/2}$, respectively. In addition, we demonstrate measurement of a near-resonance frequency electric signal with linear range of more than 91dB. This work may provide avenue for developing optically levitated nano-resonator into high-precision, continuous broadband electric field sensor.

## 2. Experimental setup

As shown in Fig. 1, the predetermined electric field is generated by applying sinusoidal voltage onto the simplest parallel plate electrodes, and the optically levitated nanoparticle placed within the electric field produces a displacement response to the field. Though this experimental apparatus of the present study is similar to that in Ref. [22], it differs in that its electrodes are composed of two horizontal steel needles that are 1 mm of gauge and placed 2.52 mm apart. This allows for producing more distinguishing changes in electric field distribution around the light field. Similar to most previously published studies, the electrodes in Ref. [23, 24] are used to calibrate the nanomechanical parameters such as particle mass and conversion factor from detection voltage to displacement, where the FDTD numerically simulated value of electric intensity is employed as a known constant. In this study, however, electric intensity generated by the electrodes is no longer a presumed parameter, but a parameter to be detected. To obtain triaxial electric intensity components at each point, an independent triaxial position detection scheme is built to obtain motion signal along each axis. The electric driving signal is then loaded onto the electrodes, meanwhile synchronously input into the phase locked loops (PLL) as a reference signal. The PLL extracts the signal components with the same frequency from the input motion signals of three axes. For stable levitation and suppression of frequency fluctuation in high vacuum, a triaxial parametric feedback scheme sums up all the feedback signals and drives a single acousto-optic modulator to cool the center of mass motion of nanoparticle.

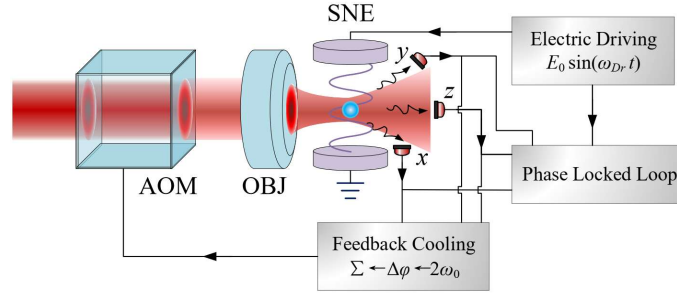

**Fig. 1.** Schematic of the experimental setup. The setup consisted of a single-beam optical trap, triaxial position detection and parametric feedback scheme, electric driving and field measurement circuit. OBJ: NA microscope objective. SNE: horizontally placed steel needles. AOM: acousto-optic modulator.

## 3. Result

### 3.1 Three-dimensional electric field vector detection

The electric intensity can be deduced from the driven displacement response and the parameter of nano-resonator. The relationship between electric field component and displacement response of the corresponding axis (taking the $x$ axis for example) is as follows:

$$E_x = \frac{2m\sqrt{S_x^{el}(\omega_{dr})\left[(\omega_{dr}^2 - \omega_x^2)^2 + \Gamma_x^2 \omega_{dr}^2\right]/\tau}}{Nq_e} \quad (1)$$

Here $m$ is the mass of the nanoparticle, $\omega_{dr}$ is the driving frequency of the electric field to be detected, $\omega_x$ and $\Gamma_x$ is the resonant frequency and damping rate of the nano-resonator, respectively. $N$ is the net charge amount of the nanoparticle, and $q_e$ is the elementary charge. Based on the motion signal with a measurement time of $\tau$, the power spectral density value $S_x^{el}(\omega_{dr})$ at the driving frequency can be extracted as the displacement response of nano-resonator. See supplement S1 for the derivation of the above formula.

We firstly moved the electrodes with a nano-positioning stage to place the particle at the symmetrical midpoint of the electrodes and measured the electric intensity at that point. As shown in Fig. 2, the normalized measured value of three orthogonal components are $\tilde{E}_x = 227.7(73)\,\text{V/m}$, $\tilde{E}_y = 15.8(14)\,\text{V/m}$, $\tilde{E}_z = 9.8(7)\,\text{V/m}$, respectively, corresponding to the case where the voltage amplitude applied at the electrodes is 1 V. The electrodes are basically capacitive and the measured equivalent impedance is about 3pF, which means that the response electric field within 1 MHz is almost frequency independent. According to the simulation result of COMSOL, three electric intensity components at this point are $\tilde{E}_{x_{th}} = 246\,\text{V/m}$, $\tilde{E}_{y_{th}} = 14.8\,\text{V/m}$, $\tilde{E}_{z_{th}} = 11.6\,\text{V/m}$, respectively, which deviate slightly from the measured values. The discrepancy between the measured and theoretical values may be a result of manufacturing error and alignment error of two steel needles, as well as the relative position error between the symmetrical midpoint and the nanoparticle.

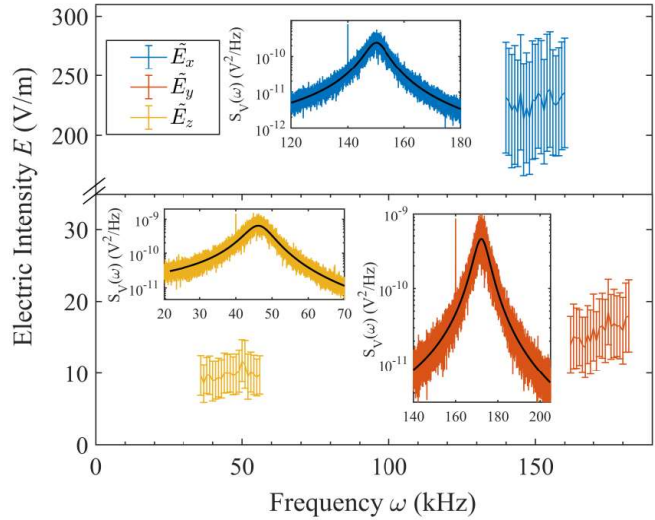

**Fig. 2.** The measured electric intensity at the symmetrical midpoint of the electrodes. A nanoparticle with a diameter of 142.8 (33) nm and charge of $N = 4$ was captured by the optical trap and the measurement was carried out at pressure of 10 mbar. Colors blue, red and yellow in the figure represent the electric field components $\tilde{E}_x$, $\tilde{E}_y$, and $\tilde{E}_z$ in three axes respectively for driving voltage of 1 V. For each component, 21 driving frequencies with intervals of 1kHz were applied, while each driving frequency was measured 100 times. The final measurement result is shown as an average of the 21 frequency points. The illustration shows the power spectral density of displacement signals in three axes, with the resonant frequencies of 150.2kHz, 172.5kHz and 46.1kHz respectively. To improve the measurement accuracy of electric intensity, the driving frequency was selected near resonance frequency.

### *3.2 Three-dimensional electric field mapping*

Taking the above position as the center point, we moved the relative positions of the nanoparticle and the electrodes along three orthogonal axes and obtained the *x* component $E_x$ of electric intensity at each point, as shown in Fig. 3(a). The variation trend of $E_x$ along each axis is consistent with the theoretical simulation results. In addition, the other two components can also be measured by using the motion signals of the other two axes in the same way and compared with the simulation results. Three-dimensional electric field mapping was realized by obtaining the resultant vectors of each component at different array points. Taking the case of *xz* plane (section y = 0), the resultant vectors of $E_x$ and $E_z$ components in this plane were measured, as shown in **Fig**. **3**(b) , together with the results of COMSOL simulation.

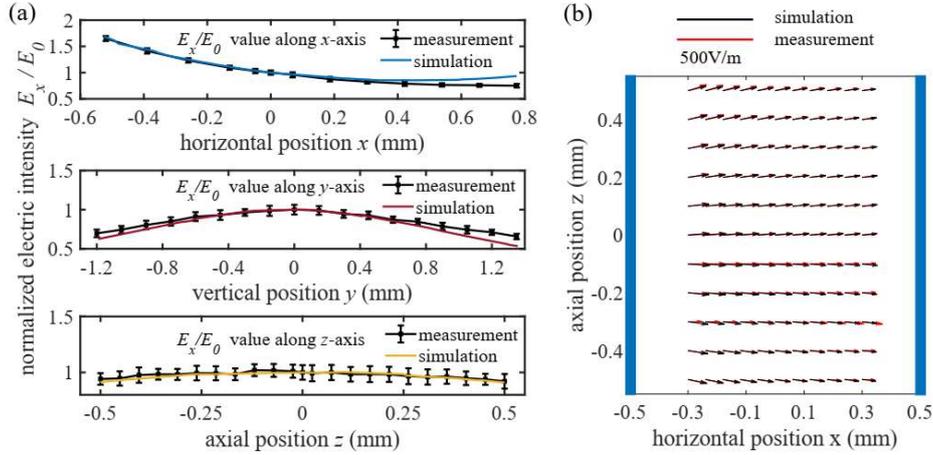

**Fig. 3.** (a) *x* component of electric intensity along the *xyz* coordinate axes. To intuitively show the variation trend, the normalized value of $E_x$ at each point is normalized by setting the *x* component $E_0$ at the center point to 1. (b) Vector field plot of $(E_x, E_z)$ in the *xz* plane, the grid spacings in the *x*-axis and *z*-axis directions are 60μm and 100μm respectively, and the blue bars on both sides represent the cross sections of steel needles constituting the electrodes.

### 3.3 Noise equivalent electric intensity

The electric intensity detection sensitivity of the nano-resonator depends on its force detection sensitivity, which can be improved by restraining thermal noise in high vacuum. But the accompanying frequency fluctuation in high vacuum would increase the complexity of model fits from the thermal noise response and the electric driven response near the resonance frequency [25]. Both issues can result in significant inaccuracies in the conversion from displacement to electric intensity with a calibrated transfer function. Therefore, feedback cooling is indispensable for realizing ultrasensitive electric field detection, although theoretically it does not improve the detection sensitivity at certain pressure condition (see Supplement S3). The displacement noise floors of nano-resonators were measured at different pressures, where the electrodes and other metal structures in chamber were grounded to isolate the residual electric field. The resulting displacement spectral densities in high vacuum for two nano-resonators with parametric feedback cooling are shown in Fig. 4(a). The fits of the displacement spectral density to the expected thermomechanical noise response superimposed on the detection noise for the nano-resonator with different feedback damping show close agreement (see Supplement S3). Beside the *x*-axis eigenmodes, the additional modes originating from crosstalk of other axes are generally visible in the thermomechanical noise response.

The thermal noise dominates over frequency range near resonance while the noise floor closely approaches the optical shot noise limit over that far from resonant. Comparing the displacement spectral density in $10^{-4}$mbar and $1.4\times10^{-7}$mbar, a reduction in gas damping, due to the balance between the thermomechanical noise and shot noise, the frequency range over which the spectral density is thermal noise limited is clearly narrowed. The displacement spectral densities in Fig. **4**(a) are converted to a noise equivalent force (NEF) by dividing the response by the theoretical transfer function of harmonic oscillator (see Supplement S3) as shown in the right side of Fig. **4**(b). Further, the noise equivalent electric intensity (NEEF) can be obtained by dividing the NEF by the charge amount as shown in the left side.

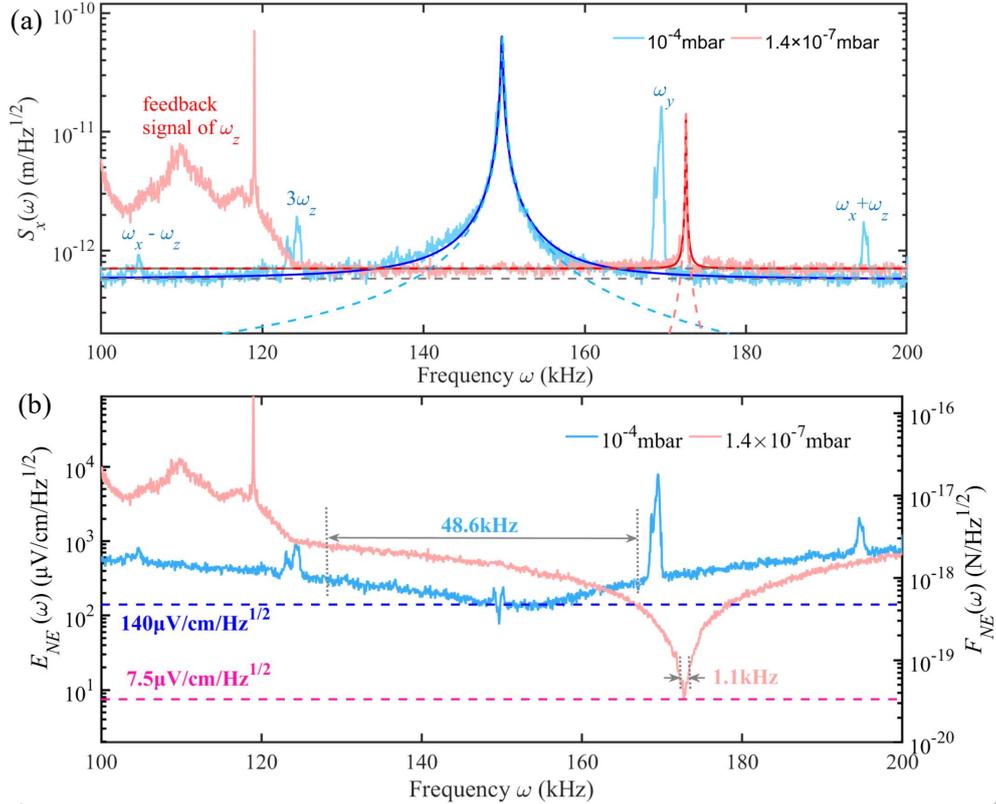

**Fig. 4.** (a) Displacement spectral densities for nano-resonators in high vacuum. Gray dashed line: detection noise. Light dashed line: fit to the thermomechanical noise model. Dark solid line: superposition of thermomechanical noise and detection noise, that is, theoretical transfer function of nano-resonator. At $10^{-4}$mbar, the cooling temperature and damping of x-axis eigenmodes were 500(6) mK and 190.4(54) Hz, respectively. The crosstalk from the other two axes brought four additional modes, as annotated in the figure. At higher vacuum, the nano-resonator was cooled to 5.1(4) mK with damping of 55.1(78) Hz. The eigenmode was increased by about 22 kHz due to the larger trapping power, making the crosstalk modes further away and not visible in the original frequency band, meanwhile, the parameter feedback signal of z-axis appeared instead. (b) The noise equivalent force (NEF) and equivalent electric intensity (NEEF). The net charge amount of the nano-resonator is set to $N = 100$. Indicated frequency bands represent the range over which the NEF is within 3dB above the thermodynamic limit (dashed lines).

As expected, the NEF and NEEF both reach the thermal noise limit near resonance frequency. When the damping is lower in higher vacuum, a lower thermodynamic limit could be provided, meanwhile, which is more difficult to reach since the thermomechanical noise must be above the shot noise. The minimum NEEF reaches the order of $1\mu V/cm/Hz^{1/2}$ at $1.4\times10^{-7}$mbar, corresponding to the case of $N = 100$, which is lower than that at $10^{-4}$mbar by more than one order of magnitude, and can be further reduced by increasing the net charge or the vacuum level. The bandwidth over which the NEEF is within 3dB above the thermodynamic limit is 48.6 kHz and 1.1 kHz for two pressures, respectively. This could be further broadened by one order of magnitude by adopting an exceptionally homodyne detection scheme and optimizing the detection noise to approach the standard quantum limit [28] (see Supplement S3).

As a comparison, the achieved minimum detectable field is superior to the reported performance using $10^4$ Rb Rydberg atoms by one order of magnitude [13], approaching the equivalent performance for an antenna dipole electronic sensor with length of 1cm [14]. One benefit of optically levitated nano-resonators is that the bandwidth of interest within which the thermal noise is above or equal to the shot noise is tunable, and the tunable bandwidth can reach the order of tens of kHz by tuning the power of trapping beam. In contrast, traditional passive dipole electronics usually need to change the structure size to achieve similar effects.

### 3.4 Linearity and linear range analysis

As a test of sensing performance for linearity and linear range of electric field sensing in *x*-axis, the nano-resonator was moved back to the center point and charged with high amount of $N = 99.0(12)$ (see Supplement S4). The measurement was performed at $5\times10^{-5}$ mbar by applying a sinusoidal electric field with frequency of 140kHz, which is 10kHz offset from the resonant to reduce the effect of frequency instability on the measurement. The driving voltage of the sinusoidal electric field generated by the steel needles was swept from 5mV to 160V, and the resulting driving responses were converted to the measured electric intensities by using Eq. (1), as shown in Fig. 5. Within the corresponding measured electric intensities ranging from 1.03V/m to 36.2kV/m, which spans over four orders of magnitude (91dB), the linearity of electric field sensing was within 10%. The minimum detectable electric field is mainly limited by the detection sensitivity of nano-resonator at $5\times10^{-5}$ mbar. While further reduction in pressure can lead to higher detection sensitivity, the accompanying instability and drift of resonance frequency become more pronounced, resulting in increased uncertainty and deviation of measured value. The maximum detectable electric field in this measurement was merely limited by the maximum output of the high-voltage amplifier (Aigtek ATA-2031), and its theoretical limit is related to the linear range of optical force and the capture region of trap [29], and ultimately limited by the response range of the detection scheme to nanoparticle displacement.

### 4. Conclusion

We have demonstrated a high-sensitivity electric field measurement technology using the optically levitated nano-resonators. By scanning the electric field distribution between parallel electrodes, the three-dimensional electric field mapping capability of the system was demonstrated. Its measuring spatial resolution depends on the motion amplitude of the nanoparticle in the equilibrium position and the manipulation accuracy of the equilibrium position, which can reach the order of nanometers. Broadband measurement at the thermodynamic limit yields a noise equivalent detection resolution of the order of $1\mu V/cm/Hz^{1/2}$ in high vacuum, which is competitive to that of previously reported electric field detection schemes. Linearity analysis near resonance shows a linear range of more than four orders of magnitude.

A higher amount of net charge is the key to further improve the detection resolution of nano-resonators. Although this can be achieved simply by using larger particle, for example, the net charge of micron-sized particle can reach the order of $10^4$ [20], the resulting force detection sensitivity is worse due to larger mass. Therefore, size of particle needs to be optimized according to these two factors to obtain the optimal electric field detection sensitivity. Although this work is based on optical levitation system, charged particles in other levitation system are eligible to be developed into a highly sensitive electric field sensor. The advantage of levitated resonator is that its resonant frequency can be adjusted from Hz to MHz according to size of particle and stiffness of potential well, to meet the application requirements of different frequency bands, especially low-frequency submarine communication.

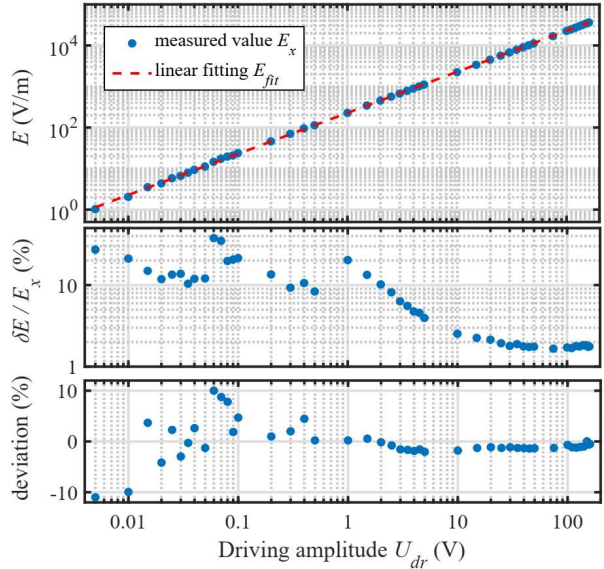

**Fig. 5.** Linearity and linear range measurement of nano-resonator. The *x*-axis component of electric field generated by the AC driving voltage was used for testing. The measurement duration of each data point is 1s. A goodness of 0.9999 was obtained by fitting the measured results with $E_{fit} = \alpha U_{dr}$ and $\alpha = 227.1 m^{-1}$. Linearity was characterized by the deviation between the measured results and the fitting values, which was calculated by $(E_x - E_{fit}) / E_{fit} \times 100\%$. The measurements for $U_{dr} > 0.5V$ and $U_{dr} \leq 0.5V$ were carried out at $3.7\times10^{-2}$ mbar and $5\times10^{-5}$ mbar, respectively.

**Funding.** This research was supported by National Natural Science Foundation of China (62005248, 62075193), Major Project of Natural Science Foundation of Zhejiang Province (LD22F050002) and Major Scientific Research Project of Zhejiang Lab (2019MB0AD01, 2022MB0AL02).

**Disclosures.** The authors declare no conflicts of interest.

**Data availability.** Data underlying the results presented in this paper are available.

**Supplemental document.** See Supplementary materials for supporting content.

*Supplementary materials*

**S1. Electric field sensing with harmonic oscillator**

A major benefit of the electric field sensor described in this work is that its dynamic response closely follows that of a one-dimensional viscously-damped harmonic oscillator, making it possible to convert from measured nano-resonator displacement to an equivalent electric intensity using a low-order model. In this section, we describe the harmonic oscillator model and the conversion between displacement and electric intensity. Much of the analysis in this section follows directly from the work of F. Ricci [1] but is specifically focused towards the optomechanical electric field sensing.

The simplified diagram of electric field sensing with optically levitated nano-resonator is described in Fig. S1, where the electric field is generated by a pair of electrodes placed in horizontal direction (x-axis direction) perpendicular to the optical axis (z-axis direction). A driving signal with amplitude of $U_{dr}$ and frequency of $\omega_{dr}$ is loaded to the electrodes, which generated a sinusoidal electric field $\mathbf{E}(t) = \mathbf{E_{dr}}\cos(\omega_{dr}t)$ near the nanoparticle. The electric field vector contained three components along orthogonal axes, written as $\mathbf{E_{dr}} = E_x\hat{\mathbf{x}} + E_y\hat{\mathbf{y}} + E_z\hat{\mathbf{z}}$. The motion signal along each axis is the driving response to the corresponding electric field component.

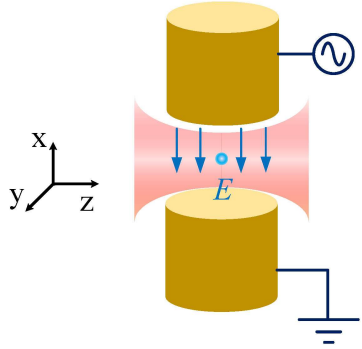

**Fig. S1.** Using optically levitated nanoparticle as nano-resonator to measure the electric field intensity near beam focus.

Taking the x-axis as an example, driven by the component $E_x\cos(\omega_{dr}t)\hat{\mathbf{x}}$, the equation of motion of the particle can be described by a thermally and harmonically driven damped resonator:

$$m\ddot{x} + m\Gamma_x\dot{x} + kx = F_{th}(t) + F_{el}(t) \tag{S1}$$

Here, $m$ is the mass of the particle, $\Gamma_x$ is the damping rate and $k = m\omega_x^2$ is the stiffness of the optical trap, with $\omega_x$ being the mechanical eigenfrequency of the oscillator. The first forcing term $F_{th}(t)$ models the random collisions with residual gas molecules in the chamber. It can be expressed as $F_{th}(t) = \sigma\eta(t)$, where $\eta(t)$ has a Gaussian probability distribution that satisfies $\langle\eta(t)\eta(t+t')\rangle = \delta(t')$, and $\sigma$ relates to the damping via the fluctuation−dissipation theorem: $\sigma = \sqrt{2k_B T_0 m\Gamma_x}$, with $k_B$ being the Boltzmann constant and $T_0$ the bath temperature. The second forcing term $F_{el}(t)$ arises from the Coulomb interaction of the charged particle with the external electric field component $E_x\cos(\omega_{dr}t)\hat{\mathbf{x}}$ and can be expressed as

$F_{el}(t) = F_{el-x}\cos(\omega_{dr}t)$, where $F_{el-x}$ is proportional to the net charge amount $N$ on nanoparticle and electric intensity $E_x$:

$$F_{el-x} = Nq_e E_x \tag{S2}$$

The power spectral density (PSD) of the harmonically driven displacement can be described as follows:

$$S_x^{el}(\omega) = S_{vx}^{el}(\omega)/c_{x/V}^2 = \frac{2F_{el-x}^2 \tau \text{sinc}^2\left[2(\omega - \omega_{dr})\tau\right]}{m^2\left[(\omega^2 - \omega_x^2)^2 + \Gamma_x^2 \omega^2\right]} \tag{S3}$$

Where, $c_{x/V}$ is the calibration factor between voltage and displacement, which can be obtained at a pressure of 10 mbar where the nanoparticle and environment are in thermal equilibrium. The mass of nanoparticle $m$ can be calculated from its radius and density and $\tau$ is the sampling time of motion signal. Therefore, the electric intensity can be obtained from the PSD at driving frequency as follows:

$$E_x = \frac{m\sqrt{S_{vx}^{el}(\omega_{dr})\left[(\omega_{dr}^2 - \omega_x^2)^2 + \Gamma_x^2 \omega_{dr}^2\right]/(2\tau)}}{c_{x/V} N q_e} \tag{S4}$$

The electric intensity generated by unit driving voltage is $\tilde{E}_x = E_x/U_{dr}$. As shown in Fig. S2, the driving frequency is selected near the resonance, and the electric intensity can be calculated with the measured amplitude $S_{vx}^{el}(\omega_{dr})$ at driving frequency by using the above formula. In previous study, F. Ricci et al proposed a simplified method to reduce measurement uncertainty, in which the electric intensity and nanoparticle mass can be converted to each other by the ratio of the electrically driven component to the thermally driven component. The premise of this method is that the thermal driven component is dominant in the PSD without driving signal, which is completely applicable in the case of high pressure and high driving frequency. However, in the case of low pressure, the noise from light source gradually dominates the PSD. In addition, for low frequency band far from the resonant frequency, environmental vibration will also introduce obvious noise component in PSD. Therefore, it is still necessary to use the measured superimposed noise rather than the ratio to obtain the electric intensity sensitivity spectrum.

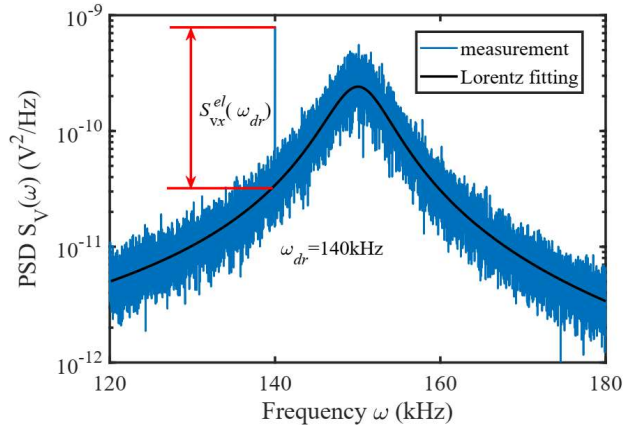

**Fig. S2.** The PSD of motion signal along *x*-axis at 10 mbar. The net charge amount on nanoparticle was $N = 4$, and a driving signal with voltage amplitude of $U_{dr} = 5$V and frequency

of $\omega_{dr}$ = 140 kHz was applied on the electrodes. The motion signal with a duration of $\tau$ = 2.72 s was obtained at a sampling rate of 1.88MHz. The resonance frequency $\omega_x$ = 150.2 kHz and damping rate $\Gamma_x$ = 8544(302) Hz were obtained by Lorentz fitting.

### S2. Error estimation of electric intensity

Obviously, reducing the uncertainty of nanoparticle mass can improve the accuracy of electric field measurement. By comparing TEM results of nanoparticle samples from different brands, nanoparticle with a nominal diameter of 150nm from Nanocym were selected in experiment, which has a relatively high particle size uniformity. As shown in Fig. S3, the mean value and deviation of particles can be obtained by averaging the size of multiple particles measured by TEM image.

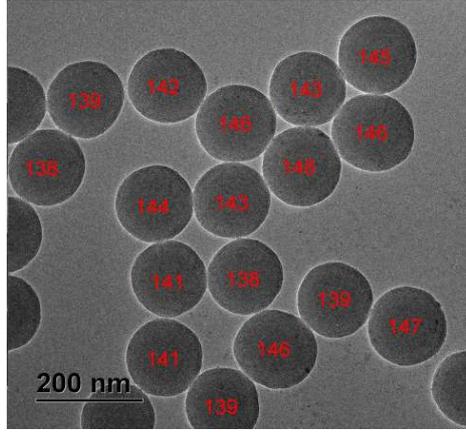

**Fig. S3.** Partial display of TEM result of particles from Nanocym. The measured diameter of each particle is indicated in the figure, and the mean value and deviation is 142.8(33) nm.

To estimate the systematic error in the calculated electric intensity, a study of all the sources of error for calculating the electric intensity component $\tilde{E}_{x_0}$ at the symmetrical midpoint must be carried out. Table S1 summarizes the absolute values and the relative uncertainties of the quantities in Eq.S4. The magnitude $S_{vx}^{el}(\omega_{dr})$ at driving frequency is the main source of error. To further improve the measurement accuracy, multiple driving frequencies were selected and the corresponding $\tilde{E}_x$ value was measured and averaged to obtain the final electric intensity value $\tilde{E}_{x_0}$ = 227.7(73) V/m.

### S3. Noise equivalent displacement and electric intensity

By extending the drive frequency to the broadband, the relationship between the harmonically driven displacement, $x_{el}(\omega)$, and driving electric intensity, $E_x(\omega)$, as a function of frequency, $\omega$, can be determined from Eq. (S1) by subtracting the Langevin force term.

$$x_{el}(\omega) = \frac{Nq_e}{m\left[(\omega_x^2 - \omega^2) + j\Gamma_x \omega\right]} E_x(\omega) = \tilde{\chi}_{el}(\omega) E_x(\omega) \quad (S5)$$

Here, $\tilde{\chi}_{el}(\omega)$ represents the transfer function between the displacement of nano-resonator and the driving electric field, which is related to the general force transfer function of the nano-resonator $\tilde{\chi}_F(\omega)$ via its net charge:

$$\tilde{\chi}_{el}(\omega) = Nq_e \tilde{\chi}_F(\omega) \tag{S6}$$

Further, similar to the force detection sensitivity of the nano-resonator, the equivalent electric intensity due to thermal noise is then:

$$E_{th} = \frac{\sqrt{2k_B T_0 m \Gamma_x}}{Nq_e} \tag{S7}$$

Obviously, $E_{th}$ is only a function of the resonator parameters ($m$, $\Gamma_x$, $T_0$ and $N$) and not a function of frequency, meaning that the thermomechanical noise floor in terms of electric intensity is flat.

**Table S1. Uncertainties Table**

| Quantity | Value | Error |
|---|---|---|
| $m$ [a] | $3.06 \times 10^{-18}$ kg | 0.0852 |
| $S_{vx}^{el}(\omega_{dr})$ | $1.298 \times 10^{-5}$ V$^2$/Hz | 0.329 |
| $\omega_{dr}$ | 140 kHz $\times 2\pi$ | 1 ppm [b] |
| $\omega_x$ | 150.2 kHz $\times 2\pi$ | < 0.001 [c] |
| $\Gamma_x$ | 8544 Hz $\times 2\pi$ | 0.0355 |
| $\tau$ | 2.72 s | 1 ppm [b] |
| $c_{x/V}$ | $9.25 \times 10^4$ V/m | 0.0561 |
| $N$ | 4 | 0 |
| $q_e$ | $1.602 \times 10^{-19}$ | $6.1 \times 10^{-9}$ |
| $U_{dr}$ | 25 V | 1 ppm [b] |
| $\tilde{E}_x$ | 231.8 V/m | 0.2012 |

[a] The particle mass was calculated from the reference density of 2.01(10) g/cm$^3$ [2] and measured diameter of $d = 142.8(33)$ nm.
[b] Nominal value from the datasheet of lock-in amplifier (Zurich Instruments MFLI).
[c] The fluctuation of resonant frequency during measurement.

When parametric feedback cooling is applied to the nano-resonator, the feedback cooling term is added to Eq. (S1), leading to an increase in the eigenfrequency and damping rate[3][4], ultimately changing the transfer function of the harmonic oscillator to

$$\tilde{\chi}_{el}^{cool}(\omega) = \frac{Nq_e}{m\left[(\omega_x + \delta\omega)^2 - \omega^2 + j(\Gamma_x + \delta\Gamma)\omega\right]} \tag{S8}$$

The equivalent electric intensity due to thermal noise with feedback cooling is rewritten as

$$E_{th}^{cool} = \frac{\sqrt{2k_B T_{cool} m(\Gamma_x + \delta\Gamma)}}{Nq_e} \tag{S9}$$

Here, $T_{cool}$ represents a lower equivalent temperature noted

$$T_{cool} = T_0 \frac{\Gamma_x}{\Gamma_x + \delta\Gamma} \tag{S10}$$

Interestingly, $E_{th} = E_{th}^{cool}$ can be deduced from Eq S9 and S10, meaning that feedback cooling does not theoretically affect the equivalent electric intensity of the nano-resonator due to thermal noise. However, in addition to thermomechanical noise, optical shot noise is the other fundamentally limiting noise source. The power spectral density of the optical shot noise is $S_{PP}=2h\nu P_a/\eta$, where $h$ is Planck's constant, $\nu$ is the optical frequency of the laser, $P_a$ is the average power reaching the photodetector, and $\eta$ is the quantum efficiency of the photodetector. This can be converted to shot noise in terms of displacement using [5]

$$x_{sn} = \frac{T_{V/i} R_{i/P}}{c_{x/V}} S_{PP}^{1/2} = \frac{T_{V/i} R_{i/P}}{c_{x/V}} \sqrt{\frac{2h\nu P_a}{\eta}} \tag{S11}$$

Here, $T_{V/i}$ and $R_{i/P}$ are the transimpedance gain and responsivity of the photodetector, while $c_{x/V}$ is the calibration factor discussed in Section S1. The shot noise in terms of electric intensity is

$$E_{sn}(\omega) = \frac{T_{V/i} R_{i/P}}{c_{x/V}} \sqrt{\frac{2h\nu P_a}{\eta}} \left| \tilde{\chi}_{el}^{cool}(\omega) \right|^{-1} \tag{S12}$$

Unlike the equivalent electric intensity in terms of thermal noise, $E_{sn}(\omega)$ is a function of frequency and gets the minimum value at the eigenfrequency of resonator. Since the thermomechanical noise and shot noise are uncorrelated, they can be summed to get the total noise equivalent displacement $x_{NE}(\omega)$ or electric intensity $E_{NE}(\omega)$. Although the optical shot noise does not represent real motion of nano-resonator, it is detection noise that analytically referred to either displacement or electric force. The best-case scenario for a nano-resonator with fixed parameters is for the thermomechanical noise to be higher than the optical shot noise, which can be done by tuning the power of trapping beam. Within the bandwidth of interest, the optical readout will measure the motion of the resonator with minimal contribution from shot noise. This is shown in Fig. S2, where the calculated noise floor is presented for a resonator with parameters similar to those described in the experiments. Three different levels of shot noise are shown. When the shot noise is reduced by one order of magnitude, the bandwidth over which the noise equivalent electric intensity is within 3dB above the thermodynamic limit increases from 370Hz to 3.6kHz by nearly one order of magnitude.

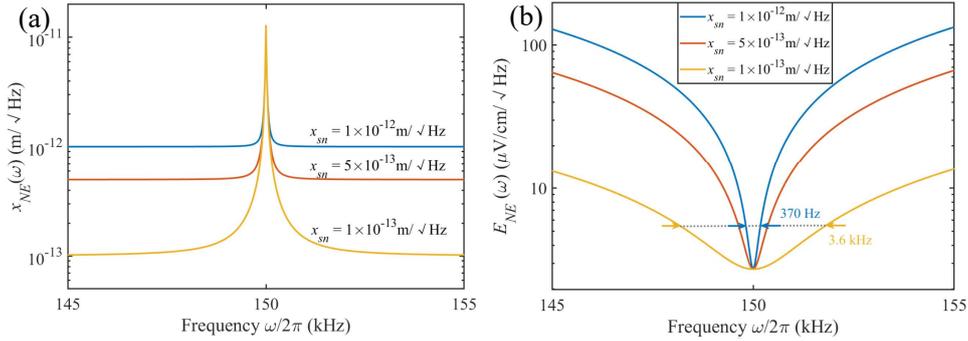

**Fig. S4.** Noise equivalent displacement and electric intensity for varying optical shot noise level. (a) Noise equivalent displacement combining thermomechanical noise and optical shot noise at three different shot noise levels. $\omega_x = 2\pi \times 150$ kHz, $\Gamma_{cool} = 16$ Hz, $m = 3\times10^{-18}$ kg, $T_{cool} = 2$ mK. (b) Noise equivalent electric intensity based on the displacement noise in (a). $N = 100$.

### S4. Controlling the net charge on nanoparticle

The charge of nanoparticle was controlled by corona discharge and Fig. S5 shows the amplitude and phase that LIA demodulated from the motion signal in $x$-axis. The motion signal was a response to the drive signal with frequency of $\omega_{dr}$ = 140 kHz. At 1 mbar, the minimum voltage step was first obtained through multiple short discharge processes and it can be considered as the voltage change corresponding to a single charge. At this pressure, the net charge of nanoparticle after multiple discharges was usually less than 10. It's observed that the net charge on nanoparticle could be raised drastically to a higher value by a long discharge process when the pressure was reduced to a range of (0.1 - 0.5 mbar). In addition, the driving voltage applied during the discharge process also affected the net charge on the nanoparticles. It seems that a higher driving voltage lead to a larger net charge as more plasma produced by ambient gas can pass through the nanoparticle. However, nanoparticle tended to enter the nonlinear region of the trap after acquiring high net charge. Therefore, when a large response signal was observed in experiment, the driving voltage should be turned off immediately to prevent nanoparticle escaping from the trap. At pressure lower than 0.1 mbar with lower gas molecular density in the chamber, the free path of electrons or ions is longer and the breakdown voltage required for discharge process is much higher. In this case, ultraviolet lamp irradiation is recommended to control the charge of nanoparticle.

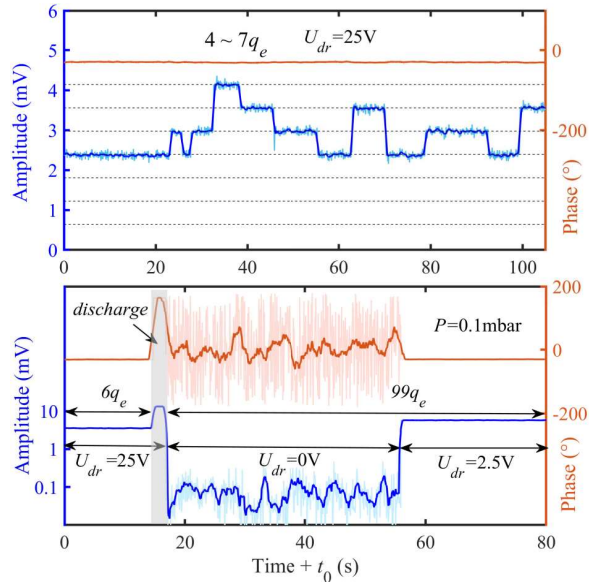

**Fig. S5.** Controlling the net charge on nanoparticle. Top: to obtain the voltage step of a single charge by multiple short discharge processes. At 1 mbar, a drive signal with an amplitude of $U_{dr1}$ = 25V was applied to the electrodes and the duration of each discharge process was about 1s. The charge of nanoparticle varied with discrete steps in the range of 4 - 7$q_e$. A single charge step of $\delta U$ = 583.7(29) μV was calculated by averaging the voltage difference value between each two adjacent steps. Bottom: to drastically increase the charge on nanoparticle by a long discharge process. The initial charge of nanoparticle was 6$q_e$ and then adjusted to a higher value after a 5s discharge process. Because the nanoparticle with high net charge entered the nonlinear region of the trap with driving signal of $U_{dr1}$ = 25V, its charge cannot be calculated directly. The driving signal was first turned off and then applied again with a smaller voltage of $U_{dr2}$ = 2.5V. The voltage step of a single charge changed to $\delta U \cdot U_{dr2} / U_{dr1}$. Using the average voltage amplitude, the corresponding nanoparticle charge was calculated as $N$ = 99.0(12).